# Design and Development of the Telescope-deployment High-vacuum teleOperated Rover (THOR) in an Airless Body Environment

Chris Womack    Miles Crist    Laura Kruger    Kelsey DeGeorge    Karynna Tuan    Jack Burns
*University of Colorado Boulder*

**Abstract**

The harsh environment on the lunar surface presents unique technological challenges for space exploration.  This paper presents research on the design and development of the Telescope-deployment High-vacuum teleOperated Rover (THOR), currently being built and tested in the Lunar and Airless Bodies Simulator (LABS) facility at the University of Colorado Boulder.  This rover is fabricated entirely out of cost-effective commercial off-the-shelf (COTS) components and materials.  THOR can potentially survive for more than one simulated year in conditions similar to that of the lunar environment, demonstrating the successful initial results of a first phase research study on material and electronic survivability in an extreme environment such as the Moon.

## Introduction

**Background:**

The LABS (Lunar and Airless Bodies) team is designing and fabricating a rover that can survive for more than a year on the lunar surface using commercial off-the-shelf parts and materials, with the first prototype being the primary discussion in this paper.  THOR can operate in a high vacuum environment and has the capability to drive through powdery, fine-grained lunar regolith simulant JSC-1a.  It also has sufficient thermal management systems to be protected against the extreme thermal variation between night and day (-150°C – 100°C).  THOR shows the success that can be achieved when using low-cost options engineered for an innovative solution.  As human exploration is trending towards relatively low cost missions as real options for exploring space, THOR is a first phase demonstration of those possibilities.

LUNAR is collaborating with Lockheed Martin to develop a lunar L2-farside exploration and science mission concept to teleoperate a rover on the lunar surface from the Orion Multi-Purpose Crew Vehicle (MPCV) orbiting the L2 (Lagrange) equilibrium point.  Astronauts would control the rover to deploy the low frequency telescope array and orchestrate a sample return from South Pole Aitken Basin [1].

**Literature Review:**

The Lunar University Network for Astrophysics Research (LUNAR) team is led by PI Dr. Jack Burns at the University of Colorado Boulder.  LUNAR proposes a revolutionary radio telescope array that will be deployed in the only proven radio-quiet environment in the inner solar system: the far side of the Moon.  This telescope will allow astronomers to study the early universe, supporting one of the top science objectives in the NASA Astrophysics Roadmap [2].  It will observe a previously unexplored period of time in our universe's history ~85-500 million years after the Big Bang using the hydrogen spin flip 21-cm line redshifted to the radio frequencies below 100 MHz [3].

To avoid the saturated radio spectrum from human generated RFI and the emission, absorption, and refraction effects of the Earth's ionosphere, the telescope will be remotely deployed on the far side of the Moon and cover many square kilometers of the lunar surface.  This antenna array will be constructed of long arms of a thin, flexible polyimide material widely known as Kapton. Kapton has been tested extensively and has a long heritage in the space industry, having been used in many space applications [4]. This material selected for the mission demonstrates stable properties over a wide temperature range, is lightweight, and is relatively inexpensive. Dipole antennae will be printed onto the thin Kapton framework material and data will be sent back to a central hub via transmission lines.

LUNAR is also collaborating with the NASA Ames Research Center Intelligent Robotics Group (IRG) to investigate how to conduct joint human-robotic missions where humans might orbit a planetary body in a capsule and control a rover on the surface. Our team and IRG conducted a simulation during which astronauts orbiting in the International Space Station remotely controlled the K10 rover on the Earth's surface, deploying an arm of the antenna array after completing only an hour of "just-in-time training". Data gathered from this will be extremely useful in learning how to conduct joint missions in the future [5] and complements the LABS team's research into low cost options for space exploration.

## Methods

### Experimental Setup:

LUNAR's Lunar and Airless Bodies Simulator (LABS) team is conducting laboratory tests in support of the goals mentioned previously by simulating airless body environments on Earth. LABS contains two thermal-vacuum chambers, each capable of achieving $10^{-7}$ to $10^{-8}$ Torr. One vacuum chamber is housed within a class 10,000 clean room and is reserved for those parts requiring a pristine, clean environment for testing. The second vacuum chamber contains a bed of JSC-1a lunar simulant regolith mined from a volcanic crater in Arizona, with an average grain size of 100 microns (Fig. 1).

An aluminum thermal table with piping to circulate nitrogen gas heated to 290°C and liquid nitrogen at -196°C through its center simulates the extreme thermal variation between lunar day (100°C) and night (-150°C). Because rapid temperature cycling is likely to create more degradation than maintaining one temperature extreme for extended durations, one year on the Moon is simulated over the length of one month, with each 24 hour period representing a lunar day or night cycle. A deuterium UV lamp is also installed in the chamber to simulate solar UV energy flux from 115-160 nm. Finally, a camera is located in the chamber to monitor the test in real time.

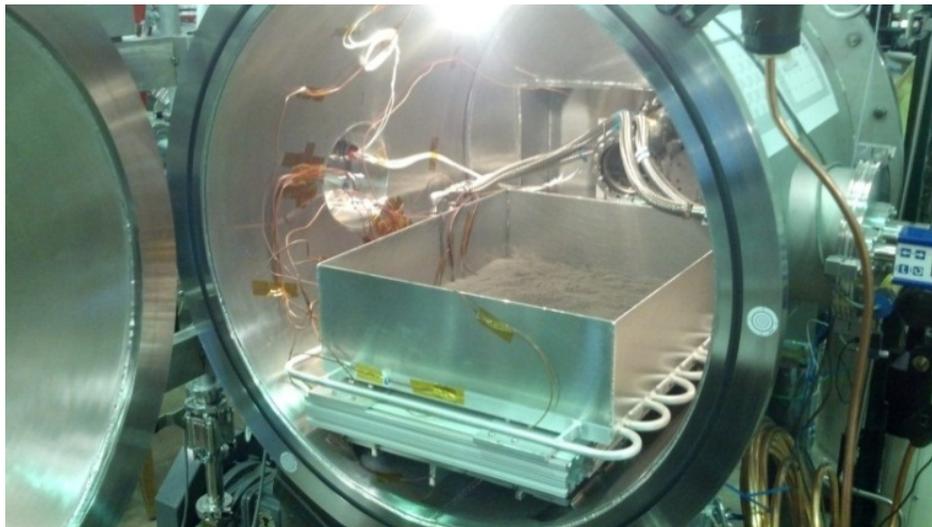

**Figure 1: The LABS thermal vacuum simulant chamber is capable of achieving high vacuum, thermal cycling between -150°C and 100°C, and provides real-time video feed.**

THOR is placed in the regolith simulant bed, initially connected to a charging dock. After vacuum is pulled, the lunar day cycle begins and the rover's first task, sent by the command station located outside of the chamber, is to deploy a simulated arm of the Kapton telescope array in a designated area. After traversing through the regolith for five minutes to demonstrate THOR's functionality, the command crew docks the rover and requests continuous temperature data be wirelessly transmitted. The first 24 hours is a lunar day cycle, holding at a temperature of 100°C. Similar to how operations will be conducted on the Moon, THOR is not operated during the night cycles and will only revive once the simulated Sun rises and the power system is recharged. Because there is no energy source to model solar recharging in the chamber, THOR receives electrical charge at the dock via copper plates and a power supply located outside of the chamber. These day and night cycles are repeated every 24 hours for one month.

**THOR Design:**

THOR's systems are specifically designed to survive for an extended period in the lunar environment with the use of inexpensive COTS components. Material selection to thermally protect onboard electronics is a key driving consideration that constrains the other systems. In order to achieve maximum traction control in the powdery regolith by allowing for wheel independence, THOR possesses all-wheel drive capabilities. The Arduino microcontroller gathers sensor data to monitor vehicle health, communicate wirelessly with the remote command center, operate an antenna deployer, and provide traverse sequence commands. The power system gives the rover the ability to operate for a minimum duration of two days and recharge during simulated lunar day.

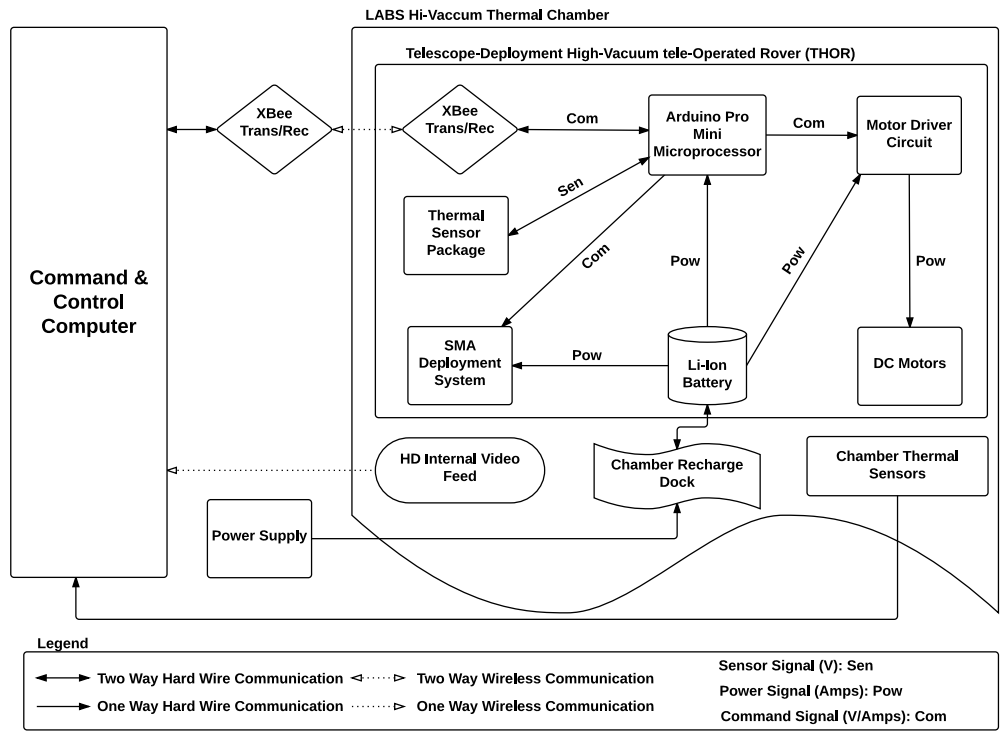

**Figure 2: Block Diagram of the entire system showing signal flow. The Arduino microcontroller redirects commands from the remote command center to all onboard systems.**

*Thermal /Structure:*

Due to the extreme environment on the Moon, thermal protection is one of the most fundamental, and constraining, requirements for THOR's structure. The rover's design is further limited by the availability of space-qualified, low outgassing materials, rated for use in a vacuum. Materials must be thermal insulators, lightweight, rigid, and cost effective. The structure must offer sufficient support for the battery and embedded systems, while minimizing thermal energy transfer.

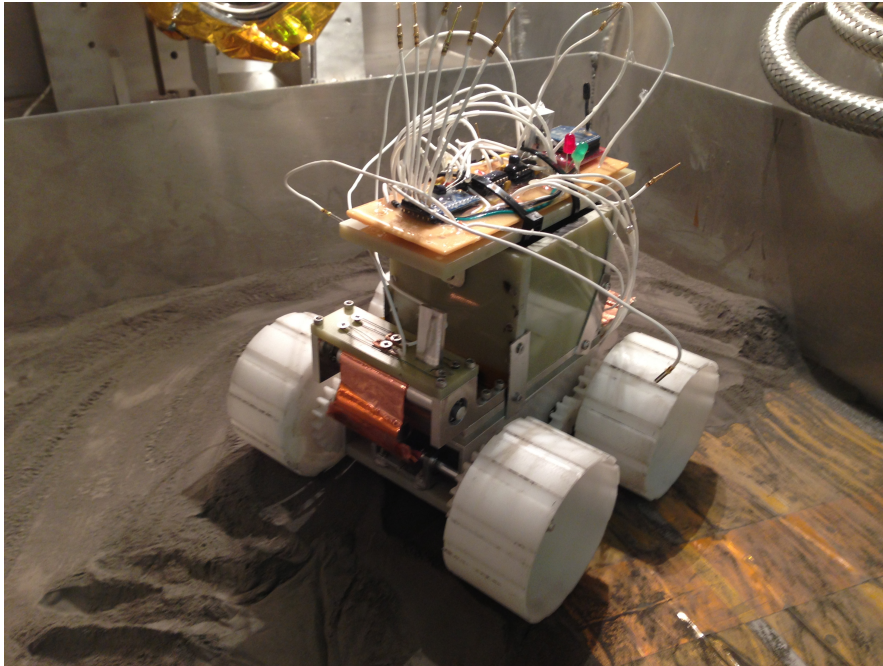

**Figure 3: T̲elescope-deployment H̲igh-vacuum tele-O̲perated R̲over (THOR) inside the lunar simulation vacuum chamber.**

The rover chassis consists of a machined aluminum 6061 U-channel as a base deck for the electrical, power, and deployment systems, as well as side fins to support the drive system (axle, motors, and gears). G-10 (garolite fiberglass epoxy laminate) insulation layers and machined Delrin (acetal resin) wheels provide additional thermal resistance to protect sensitive onboard systems.

These materials were selected for use in THOR's design based on a comparison of onboard electrical component operating temperature ranges. Given that the battery has the narrowest range (-20°C-60°C), the deck is designed to prevent the electronics from exceeding these temperature extremes. The thermal insulation materials were selected for use based on empirical methods. Samples were constructed using different layers of oxygen-free copper, aluminum, Teflon (polytetrafluoroethylene), G10, and multi-layer insulation (MLI) to determine which material combination is optimal. All of the samples performed similarly in the thermal-vacuum test, reaching a minimum temperature of $5°C$ and a maximum temperature of $47°C$. Because Teflon and G10 performed comparably, THOR's final insulation configuration is aluminum with a G10 insulation layer (Fig. 3), due to the rigidity and machinability of G10 as compared to Teflon.

To ensure survivability of the most sensitive electronic components, other thermal-vacuum tests were conducted on the battery and micro controller. These tests verified the components' ability to operate at the manufacturer-defined temperature limits, as well as the thermal structure's ability to insulate against temperatures outside of these specifications.

*Drive System:*

The main design specification for the rover's drive system is the ability to traverse through fine-grained lunar regolith. The vacuum chamber used to test THOR contains a bed of regolith simulant about 2 cm deep. A rover previously built and tested by the LABS team was only designed to operate on a flat, aluminum thermal table and would become entrenched if used in the regolith simulant. Consequently, an entirely new drive system needed to be designed for THOR. The first iteration was a two wheel chain driven system that met with little success due to the fine regolith compacting in the chain links and preventing movement.

THOR is now designed with independent all-wheel drive to provide independence of motion and high power in case of immobilization. Any wheel that becomes fixed due to miring in the regolith is able to draw the full allowable current in order to mobilize. This ability to recover a fixed wheel is crucial as losing one could mean the loss of the rover and failure of the mission.

The final component of the drive system design is the lightweight, treaded Delrin wheel (Fig. 4). Prototypes previous to the current THOR design demonstrated that wheel size is crucial. In order to avoid high centering, the rover has wide wheels to spread the rover's weight over a larger area. The wheels also have a large diameter to provide a higher angle of approach, keeping the rover from becoming trapped in mounds of regolith. Lastly, treads are necessary to create traction in the simulant.

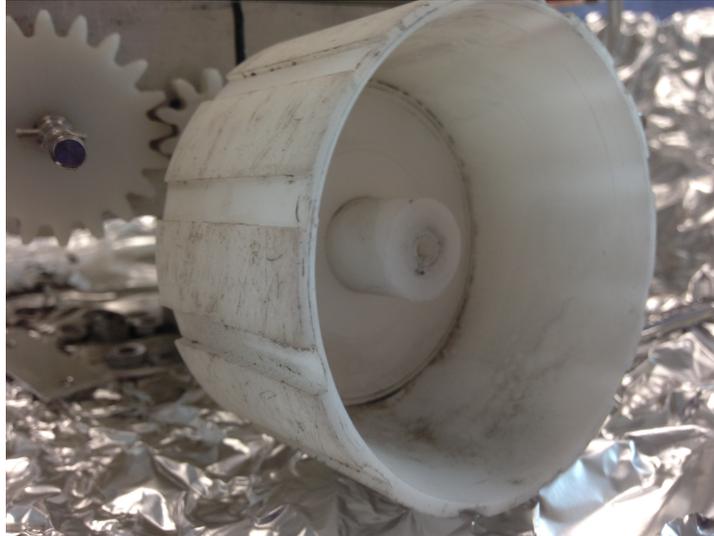

**Figure 4: Delrin wheel machined with large surface area and treading to traverse through regolith simulant.**

Highly geared DC motors are used in the design due to their large torque to weight ratio. THOR's total weight is 5.1 lbs and the total torque available is 13.8 in-lbs, yielding a torque to weight ratio of 2.7. This provides sufficient frictional force between the wheel and regolith for the rover to maneuver when stuck. The motors are suspended above a deck of G10 insulation to thermally protect the motors. Based on testing conducted with the previous iteration of THOR, the anticipated motor temperature in this design configuration was not expected to exceed the vendor operating temperature specifications of $T_{max} = 100°C$, and $T_{min} = -35°C$. A supplementary thermal-vacuum test was also conducted to ensure that prolonged exposure to ambient temperatures at $-150°C$ would not significantly deteriorate motor performance. The motors were placed in direct physical contact with the thermal table, reaching a minimum of approximately $-62°C$ after 24 hours and reviving at $-20°C$ with little readily apparent degradation, and continued to function nominally during subsequent use.

The transfer of motion from the motor to the wheels is accomplished using a direct drive system. The motors include a gearbox with a gear ratio of 67:1. From the drive axle, the motion is directed through another set of gears with a ratio of 2:1 in order to further increase the torque of the rover, resulting in a total gear ratio of 134:1. This design allows for maximum power transfer and reduced system complexity as compared to the chain drive system, which is more prone to failure.

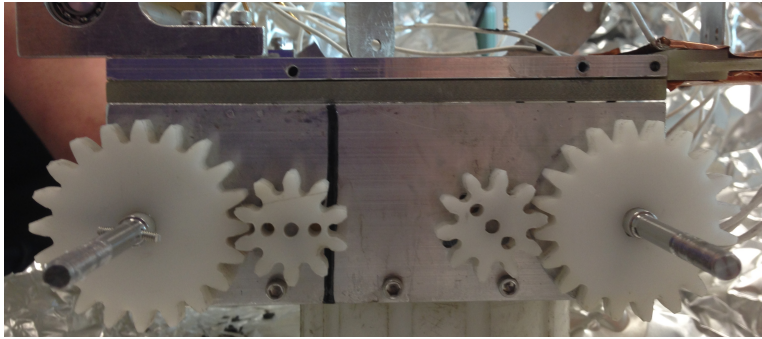

**Figure 5: Side view of THOR displaying the Delrin gears and chassis thermal structure composed of aluminum and G10.**

*Embedded Systems:*

    The embedded systems onboard THOR, as diagramed in Fig. 5, are engineered for simplicity to reduce failure risk in an extreme, remote environment where repair is not an option and diagnostics are limited. For the rover to be completely functional, this system must control four independent wheels, obtain temperature data from critical components, communicate wirelessly with a command computer, and operate an antenna deployment system. The main component of the system is an Arduino Pro Mini 5V 16 MHz microcontroller that receives and interprets commands to control rover operations. This specific model was selected because it is inexpensive, programmer friendly, and has a diverse function capability. The microcontroller sends pulse-width modulated signals and logic signals to four separate full H-bridge motor drivers. These signals dictate the speed and direction of each wheel independently. The commands for these control signals are transmitted to the microcontroller from a computer control station using a 2.4 GHz Xbee Wireless module chosen because of the interfacing simplicity with Arduino components. THOR utilizes four one-wire digital temperature sensors to gather temperature data from critical onboard components. Two of these temperature sensors are attached to the battery, one is attached to the microcontroller, and one is attached to an H-bridge motor driver. These data are transmitted wirelessly to the command computer so that the crew can ensure none of the critical components reach temperatures above the manufacturer rated values. The antenna deployment system, discussed later, is actuated via shape memory alloy wire, which contracts when a current flows through it. THOR generates this current by using a Darlington transistor array connected in the common emitter configuration activated by a logic signal from the microcontroller.

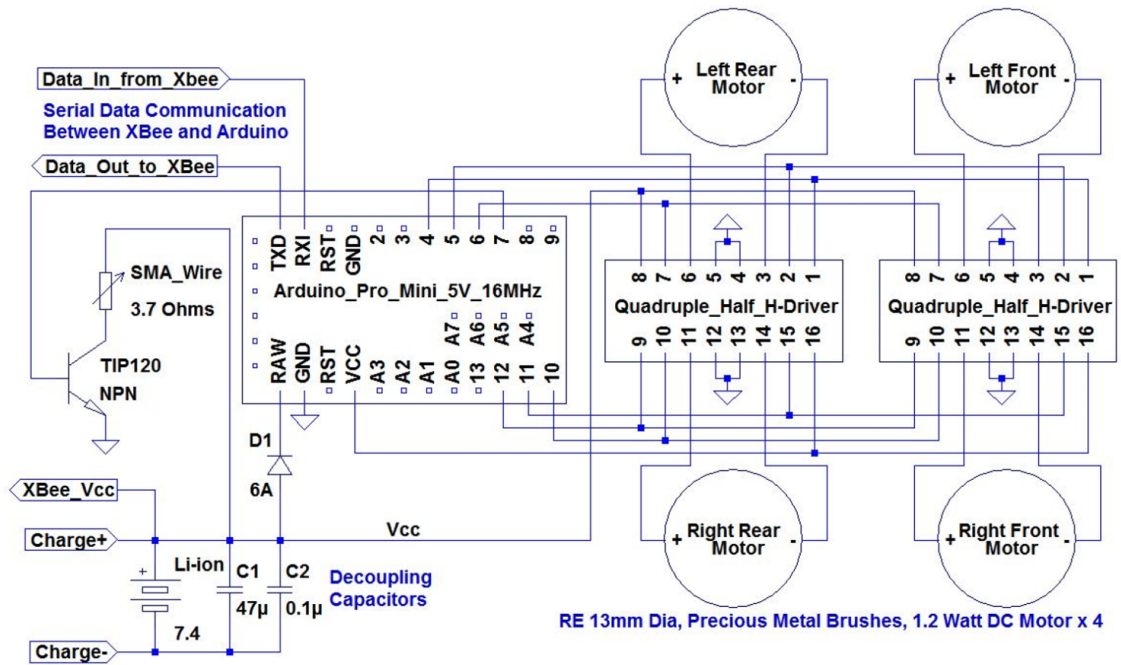

**Figure 6: Circuit Diagram of onboard electronic system for communication, processing, and control.**

*Power:*

The power onboard THOR is supplied by a 7.4V 10,400 mAh lithium ion battery. Lithium ion was selected because of its high energy density, lack of memory effect, and simple charging algorithm. The size of the battery was determined by approximating the amount of time each motor would be run and the current draw of the embedded system over a 48 hour period. This battery comes preinstalled with a protection circuit board to prevent over charging and discharging, as well as safeguarding against over-current draw so the battery cannot be damaged. The battery is potted in a thermally conductive epoxy to draw heat away from the battery and is protected from the harsh environment by a box constructed of G10. Potting also protects the battery from exploding and protects the vacuum chamber's cryopump as it is not space-rated. The battery is recharged using a simple docking station connected to a DC power supply outside the chamber. A diode is placed within the circuitry as a safety measure to prevent unanticipated reverse current, which would damage other electronic components during charging.

*Payload:*

THOR carries and deploys its antenna payload using the Kapton antenna deployment system (Fig 6). This system is designed to consume zero power until deployment. The Kapton antenna is rolled around an aluminum shaft, which is held in place by a spring loaded brake foot. A thin shape memory alloy wire is attached to a stationary surface and strung through the break foot. The length of the wire was determined based on its length contraction specifications. Upon actuation, a large current is developed in the wire, heating it and contracting it until it lifts the brake foot off the shaft. A weight on the end of the Kapton antenna arm then drops into the simulant and the antenna is unrolled passively as THOR drives forward.

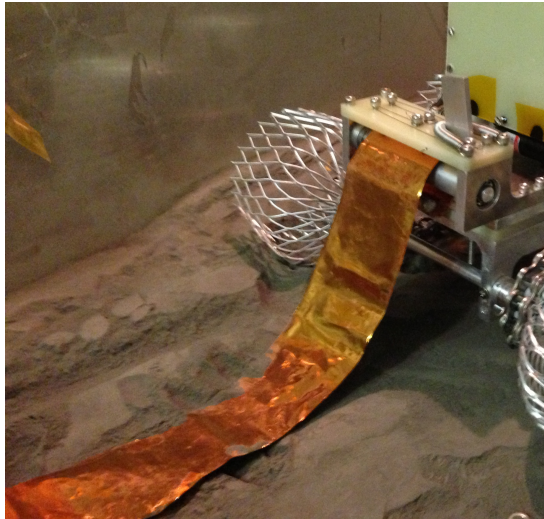

**Figure 7: A Kapton arm of the simulated telescope array is deployed from the back of THOR as it drives forward.**

## Results and Discussion

This lunar simulation successfully demonstrated that a rover built from COTS parts and readily available materials can be engineered to the extent of maintaining operations for extended exposure to lunar conditions. THOR's thermal, power, and embedded systems remained operational for the duration of the test. Each system performed according to specifications, with a few design oversights, resulting in minor loss of performance.

The most prominent issue was a temporary loss of signal (LOS) in the communications system. This was a known problem before starting the test, however it was attributed to THOR's wireless communications system operating on the same frequency (2.4 GHz) as the CU Boulder wireless internet. The LABS team tested this hypothesis empirically, and found that the LOS was more common during peak wireless traffic. Communications were always reestablished within 5-10 minutes of LOS. Because communications were not essential during peak wireless traffic hours, this was deemed an acceptable risk to proceed with testing.

THOR performed nominally until day 10.
Although the test was designed to be a 24 day simulation, LOS caused a loss in communications with THOR, forcing the LABS team to break vacuum and manually reset the embedded systems. Physically disconnecting and reconnecting the power system rectified this problem. The test resumed with revisions to accommodate a future LOS event, and no further breaks in rover communication occurred. This revision was to keep THOR docked at the charging station for the duration of the test. Upon completion of the test, THOR was successfully given a sequence of commands to ensure full functionality of all onboard systems.

An additional issue that occurred after successive testing was the failure of redundant temperature sensors attached to the battery. The reason for this failure is unknown but can be attributed to a broken solder connection or an error in the Arduino code that sends the hexadecimal algorithm to gather the temperature data. The rest of the electronic systems have remained completely functional, with the exception of the infrequent LOS.

Temperature specifications were not specifically defined for the Arduino, so the specifications for the most sensitive electronic component, the ATmega168 microprocessor ($-40°C$ to $85°C$), were used. The microprocessor experienced a temperature range from $2°C$ and $62°C$, well within the operating temperatures.

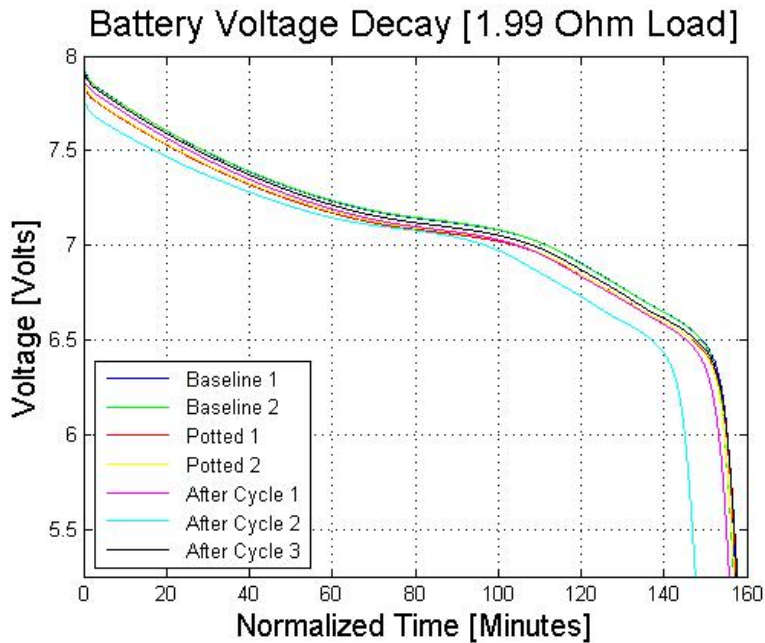

**Figure 8: Battery voltage decay showing little change between baseline readings and after potting rates. There is a noticeable variability in the decay after exposure to the lunar environment.**

The continuous function of all on board systems demonstrated that the power system performed as anticipated. The main concern with the power system was that thermal cycling, high vacuum, or potting would have some effect on voltage decay rate, thus reducing the battery capacity. Upon completion of the lunar simulation, the battery box was disassembled to investigate any battery damage caused by the thermal cycling process. The battery significantly expanded during the test, either due to thermal effects or the large pressure differential. To examine what factor may have led to this expansion, another 7.4V 10,400 mAh lithium ion battery was purchased and tested under the same experimental conditions. Baseline readings were taken before potting, after potting, and after cycling for six days to determine which treatment variable might have an effect on the voltage decay rate. These results show a possible unsystematic shift in the voltage decay rate, which can be seen in Table 1. A statistical analysis will need to be completed with a larger sample size to determine if the scattered data is an anomaly or a significant change in voltage decay rate.

**Table 1: Voltage Decay vs. Time**

| Trial | Decay Time (min) | Avg. Voltage (V) |
|---|---|---|
| Baseline | 128.34 | 7.280 |
| Baseline | 128.17 | 7.282 |
| Potted | 123.85 | 7.231 |
| Potted | 123.84 | 7.237 |
| After Cycle | 123.00 | 7.253 |
| After Cycle | 114.62 | 7.216 |
| After Cycle | 125.84 | 7.268 |

### Future Work:

While many of these systems successfully survived the duration of the simulation, this experiment is an ongoing engineering design and implementation test; thus, there will need to be improvements in future iterations of this rover to accommodate a more realistic environment. Many of the suggested improvements discussed in this section, such as drive system redundancy and the use of the least costly processor, were not applied to this rover iteration in order to reduce the overall complexity. Other improvements, such as inductance insulation and thermal protection, will be natural progressions, as the test environment becomes more realistic. Lastly the payload deployment system (PDS) design will be improved for better operation.

*Redundancy:*

The main redundancy that will be implemented on the next design iteration of this rover will be the drive system. The current system uses four wheels with four independent motors. While the independence between the motors allows for the others to continue operating when one fails, the four-point contact of the rover with the simulant creates a one-step failure if a single motor should become non-operational. To remedy this flaw, adding another set of wheels with independent motors will allow the rover to continue to operate, though impaired, allowing the test or mission to be salvaged. In future iterations, being able to lift any wheel in the event of a failure would allow continued unimpaired operation of the rover. This is a key component to many of the current rovers in operation today, and will need to be added to THOR in order for realistic operation and testing to be viable.

*Inductance:*

THOR's design did not take into account, nor mitigate for, the possibility of electrical interference between wires [6], with the wires to the motors and two of the temperature sensors being longer than necessary (20 cm). After the simulation, testing was conducted to find the root cause of the LOS problems and inductive interference was investigated as a possible cause; however, no conclusive results were obtained. For future designs, actions will be taken to decrease the chance of any inductive interference. All high current carrying wires will be twisted together and covered in a copper or aluminum shield connected to a common ground. Wire lengths will be as short as possible and any signal wires will be routed through areas that do not contain high power wires.

*Wireless communication:*

With the cause of the LOS still irreproducible, future iterations will be outfitted with higher quality wireless communication modules. The wireless system needs a significant increase in signal strength. The system should also be operated on a different frequency band than the 2.4 GHz at which it is currently operating in order to avoid interference with other wireless communication systems. Channel hopping can also be introduced to avoid interference with other communications operating in the lab area.

*Thermal Protection:*

Thermal insulation modifications will be necessitated by a change in the simulated environment. Currently, the thermal table provides the temperature variability from below rather than the more realistic radiation from above. This simplifies the analysis and design of THOR in terms of where to place the insulation and how much is needed. However, to create a more realistic environment, a thermal shroud will be added to the vacuum chamber to simulate radiant heating and cooling. This will increase the overall rate of heat transfer, as well as require three dimensional thermal energy analysis. The next rover iteration will need a new thermal insulation design that will enclose the onboard embedded systems currently located on the top of THOR, as well as an active heating and cooling system. Several options are currently being explored, but it will require further testing and prototyping for a final decision to be made. One innovative and intriguing option is to bury the electronics package in the regolith to reduce the rate of thermal transfer to the electronics. As mentioned before, lunar regolith has a low thermal conductivity and would shield components from extreme temperature variations. This would minimize the temperature fluctuations of the electronics reducing the amount of power needed to keep the electronics package within operational temperatures.

*Payload Deployment Systems:*

The payload deployment system was modeled after the NASA Ames K10 Rover's antenna deployment system used in the 2013 ISS telerobotics simulations. While this design is functional and simple, it lacks certain capabilities. The system uses a servomotor to control a brake foot on the shaft that holds the Kapton film. This limits the system to only an "on/off" control method. Due to material and budget constraints, the deployment system design for THOR could not have the same control method as the ISS simulation. The deployment system could only lift the brake foot once, leaving it unable to stop the deployment if an error occurred. Future iterations of the deployment system will use an active control method, ideally, a motor controlling the rotation of the shaft directly. This would allow complete control over the speed of deployment and give the rover the capability to roll the antenna back up if something went wrong.

As the simulation's realism increases, the deployment system should be designed with the capability to hold multiple antenna arms so that a new one can be loaded into the deployment system after one has been deployed, mimicking a rover depositing multiple antenna array arms on the Moon.

## Conclusion

The Global Exploration Roadmap identified planetary surface telepresence as a key component of the larger strategy for solar system exploration [7]. As human exploration becomes increasingly frequent and commercialized, it is essential that cost-effective and reproducible technology options are generated. It is also important to understand how to conduct human-guided remote operations from a control station. THOR demonstrates the feasibility of engineering a rover that is capable of surviving extreme environmental conditions, using only widely available, cost-effective COTS materials. Future work in preparation of LUNAR's antenna array deployment mission will continue at LABS to improve this initial proof-of-concept rover design and control, complementing the telerobotics simulations at NASA Ames.

## About Student Authors

The LABS Team consisted of three student employees and two student volunteers. The three student employees, Laura Kruger, Chris Womack, and Miles Crist, oversaw the design and implementation of all onboard systems. Laura Kruger acted as Lab Manager and oversaw the production, test planning, and procurement of materials. She also assisted in designing the thermal structure, deployment system, and mobility system. Chris Womack oversaw software development, thermal structure design, and mobility design. Miles Crist managed the embedded system design, deployment system, communications, fabrication, as well as aiding in the design of the thermal structure and mobility system. The two student volunteers, Kelsey DeGeorge and Karynna Tuan aided in testing, vacuum chamber procedures, and early prototyping.

Acknowledgements
The LABS team would like to recognize the support provided by the NASA Lunar Science Institute Cooperative Agreement NNA09DB30A. We would also like to acknowledge Dr. Joe Lazio (JPL), Dr. Robert MacDowall (NASA's Goddard Space Flight Center), Dr. Issa Nesnas (JPL), and Dr. Terry Fong and Maria Bualat at NASA Ames Research Center Intelligent Robotics Group (IRG).

Student Author's Bio

Chris Womack is currently pursuing a B.S. in both Mechanical Engineering, and Electrical and Computer Engineering. He has been with the LUNAR team for 2 years developing a tele-operated rover. Currently he is collaborating with Lockheed Martin to extend this application to remote control via web apps for telerobotic operations and testing.

Miles Crist is a mechanical engineer who recently graduated from the University of Colorado at Boulder. He has worked for NASA's Lunar Science Institute conducting material and electronic research in support of a lunar far side radio telescope array. Miles operated and maintained a thermal vacuum chamber and was part of a team that developed the Telescope Deployment High Vacuum teleOperated Rover (THOR). Miles now works for Crane Aerospace and Electronics designing fuel pumps for a large variety of aircraft.

Laura Kruger is a graduate student at the University of Colorado Boulder. She graduated with a BA in Astrophysics and is currently pursuing a MS in Aerospace Engineering and ME in Engineering Management. After working with the LUNAR team for five years as laboratory manager for CU's lunar simulation thermal-vacuum facility, Laura joined Ball Aerospace as a Systems Engineer in August of 2014.

Kelsey DeGeorge is a recent graduate receiving her Bachelor of Science in Mechanical Engineering from the University of Colorado at Boulder. She previously worked as an intern for Shear Engineering Corporation and Real D, Inc. She is currently a Manufacturing Engineer for Schlumberger in Houston, TX providing solutions for both design and manufacture of isolation valves for oilfield services.

Karynna Tuan is currently working towards her graduate degree in the 5-year MS/BS program, majoring in Aerospace Engineering with a focus area in Bioastronautics, the study and support of human spaceflight at CU Boulder. She has worked at Space Grant with the PolarCube team for a year, developing a CubeSat to study the Earth's tropospheric weather patterns and monitor and record temperature profiles. She is continuing her work experience in the industry field at Sierra Nevada Corporation, Space Systems as a Manufacturing Intern.

Press                                                                                                      Summary
Joint human-robotic space exploration involving the teleoperation of rovers on planetary surfaces is a key strategy defined by the ISECG's Global Exploration Roadmap. In support of this goal, the LABS (Lunar and Airless Body Simulator) team at the University of Colorado Boulder has developed a lunar simulation thermal-vacuum facility to create an analog environment in which to conduct survivability testing. LABS built a rover, the Telescope-deployment High Vacuum tele-Operated Rover, that can operate for extended periods of time in the harsh lunar environment. The rover is fabricated entirely out of cost-effective commercial off-the-shelf (COTS) components and materials, showing the potential of practical, low cost options for future space exploration.